\documentclass[aps,prd,10pt]{revtex4-1}
\usepackage{amsmath}
\usepackage{tabularx}

\newcolumntype{Q}[1]{>{\centering\arraybackslash}p{#1}}
\begin{document}
 
\title{Analytical results for group averaged scattering cross
  sections of high temperature plasmas}
\author{Jens \surname{Fiedler}}
\email{jens@stop901.de}
\affiliation{Fraunhofer Institute for Technological Trend Analysis, 
  P.O. Box 14 91, 53864 Euskirchen, Germany}

\begin{abstract}
  Multi-group method is an accepted technique for approximately
  solving the equation of radiative transfer. In this paper, group
  averaged transfer scattering cross sections, required for solving
  the equation of radiative transfer in a multi-group approach, are
  presented. Compton scattering is approximately described by the
  Compton cross section for free electrons at rest. In the low photon
  energy limit analytical results for scattering coefficients have
  been derived.
\end{abstract}
\maketitle

\section{Introduction}
Thermal radiation pressure, energy density and flux are determined by
the specific intensity which is the solution of the equation of
radiative transfer (ERT) \cite{Pomraning:73}. Solving the full
equation of radiative transfer is a complex and cost intensive
task. Several methods of solving the ERT approximately have been
studied in literature. In regions where the mean free path of photons
is small compared to the dimension of the system, diffusion techniques
are used widely. These methods have been simplify further by taking
overall frequency spectra integrated quantities into account
\cite{Olson:00}. To effectively reduce technical effort, diffusion
methods combined with flux limiters have been applied to optical thin
systems too. Morel \cite{Morel:00} showed that the accuracy of such
methods is of zeroth and first order compared to the exact transfer
asymptotics. To enhance accuracy while solving the ERT one applies the
multi-group method. This method divides the frequency spectrum into
several frequency groups.  All photons belonging to one frequency
group have the same frequency-independent averaged properties that are
characteristic for that group. This method is at least accurate for
coarse frequency groups. The purpose of this paper is to derive
analytically multi-group scattering and transfer scattering cross
section in low photon energy limit. In the first part the derivation
of the multi-group ERT is repeated. Based on that results various
analytical opacity functions for the scattering contributions are
found. It is shown, that all appearing expressions for the scattering
opacities can be reduced to the problem of solving a special integral
type. The results are compared with early six-group results obtained
numerically by Pritzer. The extension of the achieved opacity
functions to fine structured frequency groups is straightforward.

\section{Multi-group method}

\subsection{Group integrated quantities}

The radiation transport equation neglecting induced scattering terms
reads \cite{Pomraning:73}

\begin{equation}\label{eqn.ERTwithoutInducedScattering}
  \frac{1}{c}\frac{\partial I_{\nu}(\mathbf{\Omega})}{\partial t} +
  \mathbf{\Omega}\cdot\nabla I_{\nu}(\mathbf{\Omega})
  = \Sigma'_a(\nu)\left(B_{\nu}(\theta) - I_{\nu}(\mathbf{\Omega})\right) - 
  \Sigma_s(\nu)I_{\nu}(\mathbf{\Omega})
  +\int^{\infty}_0 d\nu'\;\int_{4\pi} d\mathbf{\Omega'}\; \frac{\nu}{\nu'}
  \Sigma_s(\nu'\rightarrow\nu,\mathbf{\Omega}'
  \cdot\mathbf{\Omega})I_{\nu'}(\mathbf{\Omega'}).
\end{equation}

$I_{\nu}\equiv I_{\nu}(\mathbf{r},\mathbf{\Omega},t)$ is the specific
intensity, $c$ the speed of light, $t$ the time, $\nu$ the photon
frequency, $\Sigma'_a$ the macroscopic absorption coefficient
corrected for induced emission effects, $\theta = k_B T$ the kinetic
temperature, $k_B$ the Boltzmann constant, $T$ the temperature,
$\mathbf{\Omega}$ the direction of flight of the photons and
$\Sigma_s$ the macroscopic scattering coefficient. The source function
is of Planckian type and given by $B_{\nu}(\theta)$

\begin{equation}\label{eqn.PlanckFunction}
 B_{\nu}(\theta) = \frac{2h\nu^3}{c^2}\left(\exp\left(h\nu/\theta\right) - 1\right)^{-1}.
\end{equation}

$h$ is the Planck constant. The spatial and time dependencies in
$I_{\nu}$ are suppressed.  Following Pritzker
et~al. \cite{Pritzker:76} the photon spectrum is divided into six
energy groups.  The energy groups are presented in table
(\ref{Table.PhotonEnergy6Groups}). The group identifier is $g$.

\begin{table}[htbp]
 \begin{center}
  \begin{tabular}{|Q{10mm}|Q{13mm}|Q{13mm}|Q{13mm}|Q{13mm}|Q{13mm}|Q{13mm}|}
   \hline
   g & 1 & 2 & 3 &  4 &  5 & 6\\
   \hline
   \hline
   $h\nu_g$     & 5.00e+02 & 6.40e+01 & 1.60e+01 & 4.00e+00 & 1.00e+00 & 2.50e-01\\
   \hline
   $h\nu_{g+1}$ & 6.40e+01 & 1.60e+01 & 4.00e+00 & 1.00e+00 & 2.50e-01 & 1.00e-04\\
   \hline
   \end{tabular}
   \caption[Photon energy groups.]{Photon energy groups. The photon
     energy is given in keV. $h\nu_g$ means the upper photon energy
     boundary and $h\nu_{g+1}$ means the lower photon energy
     boundary.}
  \label{Table.PhotonEnergy6Groups}
 \end{center}
\end{table}

The integration of (\ref{eqn.ERTwithoutInducedScattering}) over the
energy groups leads to

\begin{equation}\label{eqn.MERT}
 \left(\frac{1}{c}\frac{\partial}{\partial t} +
 \mathbf{\Omega}\cdot\nabla\right)I_g(\mathbf{\Omega}) =
 -\left(\Sigma'_{ag} + \Sigma_{stg}\right)I_g(\mathbf{\Omega}) + Q_g +
 \int_g d\nu\;\mathcal{C}\\
\end{equation}

where $\displaystyle \int_g d\nu\;[\dots]$ is a shortened term for
$\displaystyle \int^{\nu_g}_{\nu_{g+1}} d\nu\;[\dots]$. $\mathcal{C}$
indicates the in-scattering contribution

\begin{equation}
 \mathcal{C} = \int^{\infty}_0 d\nu'\;\int_{4\pi} d\mathbf{\Omega'}\; 
    \frac{\nu}{\nu'}\Sigma_s(\nu'\rightarrow\nu,\mathbf{\Omega}'
    \cdot\mathbf{\Omega})I_{\nu'}(\mathbf{\Omega'})
\end{equation}

and will be considered later. The variables $\Sigma'_{ag}$, $I_g$,
$B_g$, $Q_g$, $\Sigma_{stg}$ denote the frequency group (see table
(\ref{Table.PhotonEnergy6Groups})) integrated quantities of the
macroscopic absorption coefficient, specific intensity, black body
radiation, emission source and the macroscopic scattering coefficient,
respectively. Formally, those quantities are given by
\cite{Pomraning:73}

\begin{eqnarray}
  I_g\left(\mathbf{\Omega}\right) &=& \int_g
  d\nu\;I_{\nu}\left(\mathbf{\Omega}\right)\\
  \Sigma'_{ag} &=& \left(\int_g d\nu\;
  \Sigma'_a\left(B_{\nu}(\theta) - I_{\nu}\right)\right)\times
  \left(\int_g d\nu\;\left(B_{\nu}(\theta) - I_{\nu}\right)\right)^{-1}\\
  \Sigma_{stg} &=& \frac{1}{I_g}\int_g d\nu\; \Sigma_s(\nu)I_{\nu}\\
  Q_g &=& \int_g d\nu\;\Sigma'_a(\nu)B_{\nu}(\theta)\\
  B_{g}(\theta) &=& \int_g d\nu\; B_{\nu}(\theta)\label{eqn.GroupPlanck}.
\end{eqnarray}

In a local thermal equilibrium regime $\Sigma'_{ag}$ and $\Sigma_{stg}$ can be
approximated by their Planck averaged quantities. In this case $\Sigma'_{ag}$
and $\Sigma_{stg}$ read

\begin{eqnarray}
  \Sigma'_{ag} &=& \frac{1}{B_g}\int_g d\nu\;\Sigma'_a(\nu)B_{\nu}(\theta)\\
  \Sigma_{stg} &=& \frac{1}{B_g}\int_g d\nu\;\Sigma_s(\nu)B_{\nu}
 (\theta)\label{eqn.SigmaSTG}.
\end{eqnarray}

Generally, the radiation source is given by the ratio of the emission and
absorption coefficient. In local thermal equilibrium the source is of
Planckian type. Therefore one has for the group emission source

\begin{equation}\label{eqn.GroupAbsorption}
  Q_g = \int_g d\nu\;\Sigma'_a(\nu)B_{\nu}(\theta) = \Sigma'_{ag} B_g.
\end{equation}

The coefficients $\Sigma'_{ag}$ and $Q_g$ are not considered within this
report. These contributions are discussed widely in \cite{Pritzker:76} and
\cite{Li:09}.


\subsection{Group integrated Planck function}
Inserting the Planck function in (\ref{eqn.GroupPlanck}) leads to

\begin{equation}
  B_g = \frac{2\theta^4}{c^2 h^3}
  \int^{u_g}_{u_{g+1}} du\;u^3 \left(\exp(u) - 1\right)^{-1}
\end{equation}

where $u_g$ and $u$ are defined as $u_g=h\nu_g/\theta$ and
$u=h\nu/\theta$. In contrast to the numerical evaluation of
(\ref{eqn.GroupPlanck}) in \cite{Pritzker:76} an analytical solution
of this integral is possible. A discussion of this integral can be
found in the appendix. Solving the integral yields

\begin{equation}
 B_g = \frac{2\theta^4}{c^2h^3}\mathcal{I}_3\left(u_{g+1},u_g\right).
\end{equation}

The results  for the six group Planck spectrum $B_g$ in units of keV/cm$^2$s
are presented in table (\ref{Table.SixGroupPlanckSpectrum}).

\begin{table}[htbp]
  \begin{center}
    \begin{tabular}{|Q{8mm}|Q{1.7cm}|Q{1.5cm}|Q{1.5cm}|Q{1.5cm}|Q{1.5cm}|}
      \hline
      g/$\theta$ & 1.00e-01 & 3.16e-01 & 1.00e+00 & 3.16e+00 & 1.00e+01\\
      \hline
      \hline
      1 & 9.329e-243 & 2.911e-52 & 1.387e+09 & 4.852e+28 & 2.246e+35\\
      \hline
      2 & 4.277e-36 & 4.428e+12 & 1.758e+28 & 4.834e+33 & 1.593e+36\\
      \hline
      3 & 9.228e+14 & 2.585e+27 & 8.230e+31 & 1.426e+34 & 2.194e+35\\
      \hline
      4 & 1.951e+26 & 1.161e+30 & 1.149e+32 & 1.251e+33 & 5.657e+33\\
      \hline
      5 & 1.443e+28 & 8.352e+29 & 6.923e+30 & 2.887e+31 & 9.936e+31\\
      \hline
      6 & 5.801e+27 & 3.802e+28 & 1.490e+29 & 5.025e+29 & 1.623e+30\\
      \hline
    \end{tabular}
    \caption[Six-group Planck spectrum]{Six-group Planck spectrum
      $B_g$ in units of $\mbox{keV}/\mbox{cm}^2\mbox{s}$. $g$ is the
      group index and $\theta$ is the temperature $\mbox{keV}$. No
      contributions to $B_g$ are given for high energy groups and
      small temperatures. In that case Pritzker et~al.
      \cite{Pritzker:76} used a transformed presentation for
      $B_{\nu}(\theta)$ called $B^*_{\nu}(\theta)$. To avoid numerical
      problems while evaluating group constants $B^*_g$ is defined by
      \mbox{$B^*_{\nu}(\theta)=\exp(u_{g+1})B_{\nu}(\theta)$}.  In the
      present analytical considerations no such transformations are
      necessary.}
    \label{Table.SixGroupPlanckSpectrum}
  \end{center}
\end{table}

\subsection{Scattering cross section}

The scattering contribution appearing in (\ref{eqn.MERT}) is obtained by
using the angle integrated Compton cross section \cite{Pomraning:73}

\begin{equation}\label{eqn.KleinNishinaAngleIntegated}
  \Sigma_s(\nu) =
  \frac{3}{4}\Sigma_{Th}\left\lbrace\left(\frac{1+\gamma}{\gamma^3}\right)
  \left[\frac{2\gamma(1+\gamma)}
  {1+2\gamma} - \ln(1+2\gamma)\right] +\right.
  +\left.\frac{1}{2\gamma}\ln(1+2\gamma) - \frac{1+3\gamma}{(1+2\gamma)^2}\right\rbrace 
\end{equation}

where $\Sigma_{Th}=0.665\times 10^{-24}\;n_e$ is the macroscopic
Thomson cross section, $n_e$ the electron density and $\gamma=h\nu/m_e
c^2$. Using (\ref{eqn.SigmaSTG}) and the abbreviation \mbox{$\gamma_g
  = h\nu_g/m_e c^2$} one obtains the Planck averaged scattering
contribution

\begin{equation}\label{eqn.Stg}
 \begin{split}
  \Sigma_{stg} &= \frac{\Sigma_{Th}}{B_g}\left(\frac{3(m_ec^2)^4}{2c^2 h^3}\right)
  \int^{\gamma_g}_{\gamma_{g+1}}\frac{d\gamma\gamma^3}{\left(\exp(\gamma m_e
 c^2/\theta)-1\right)}\left\lbrace\left(\frac{1+\gamma}{\gamma^3}\right)\times\right.\\
  &\left.\times\left[\frac{2\gamma(1+\gamma)}{1+2\gamma} -
  \ln(1+2\gamma)\right] + \frac{1}{2\gamma}\ln(1+2\gamma) - 
 \frac{1+3\gamma}{(1+2\gamma)^2}\right\rbrace
 \end{split}.
\end{equation}

The integration is carried out numerically. The results are given in table 
(\ref{Table.SixGroupPlanckMeanComptonCS}). In case of small photon energies,
e.g. $\gamma\ll 1$, equation (\ref{eqn.Stg}) is integrated analytically. The
scattering cross section (\ref{eqn.KleinNishinaAngleIntegated}) corrected to
second order is \cite{Pomraning:73}

\begin{equation}
 \Sigma_s = \Sigma_{Th}\left(1-2\gamma+\frac{26}{5}\gamma^2\right).
\end{equation}

Inserting the above expression in (\ref{eqn.SigmaSTG}) yields

\begin{equation}
 \begin{split}
  \Sigma_{stg} &= \frac{\Sigma_{Th}}{B_g}\left(\frac{2(m_ec^2)^4}{c^2 h^3}
  \right)\left\lbrace \int^{\gamma_g}_{\gamma_{g+1}}\frac{d\gamma\gamma^3}
  {\left(\exp(\gamma m_e c^2/\theta)-1\right)}\right.\\
 &\left.-2\int^{\gamma_g}_{\gamma_{g+1}} \frac{d\gamma\gamma^4}{\left(\exp(\gamma m_e
 c^2/\theta)-1\right)}+\frac{26}{5}\int^{\gamma_g}_{\gamma_{g+1}}
  \frac{d\gamma\gamma^5}{\left(\exp(\gamma m_e
c^2/\theta)-1\right)}\right\rbrace. 
 \end{split}
\end{equation}

$\Sigma_{stg}$ is related to

\begin{equation}\label{eqn.ScatteringSigmaSTGAnalytic}
 \Sigma_{stg} = \frac{\Sigma_{Th}}{B_g}\left(\frac{2\theta^4}{c^2 h^3}\right)\left\lbrace
 \mathcal{I}_3(u_{g+1},u_g)\right.
 \left.-2\alpha\mathcal{I}_4(u_{g+1},u_g)+\frac{26}{5}\alpha^2
 \mathcal{I}_5(u_{g+1},u_g)\right\rbrace
\end{equation}

where $\alpha$ and $u_g$ are given by $\displaystyle \alpha=\theta/m_e c^2$
and $u_g=\gamma_g/\alpha$. The results of expression
(\ref{eqn.ScatteringSigmaSTGAnalytic}) are presented in table 
(\ref{Table.SixGroupPlanckMeanComptonCSAnalytic}).


\subsection{Scattering transfer cross section}\label{sec.Scatt}

The scattering part $\mathcal{C}$ of (\ref{eqn.ERTwithoutInducedScattering}) reads

\begin{equation}
  \mathcal{C}\left[I_\nu\right] = \int_{4\pi} d\mathbf{\Omega'}\;\int^{\infty}_{0} d\nu'
  \frac{\nu}{\nu'}\Sigma_s(\nu'\rightarrow\nu,
  \mathbf{\Omega}\cdot\mathbf{\Omega'})I_{\nu'}(\mathbf{\Omega'}).
\end{equation}

The integration of $\mathcal{C}\left[I_{\nu}\right]$ over $\nu$ by splitting the
integrals in different photon energy intervals gives the frequency averaged
in-scattering contribution

\begin{equation}
  \mathcal{C}_g\left[I_{\nu}\right] = \int_{4\pi} d\mathbf{\Omega'}\;
  \sum^{gmax(g)}_{g'=gmin(g)}S_{g'g}
\end{equation}

where $S_{g'g}$ is defined by

\begin{equation}
  S_{g'g} = \frac{1}{I_{g'}}\int_{g'} d\nu'\;\int_g d\nu\;
  \frac{\nu}{\nu'}\Sigma_{s}(\nu'\rightarrow\nu,\mathbf{\Omega'}\cdot\mathbf{\Omega}) 
  I_{\nu'}(\mathbf{\Omega'}).
\end{equation}

\texttt{gmin(g)} indicates the smallest and \texttt{gmax(g)} the highest
energy group $g'$ from which a photon is able to scatter into group $g$. By
the summation over $g'$ all possible photon scattering contributions from the
energy group $g'$ into the energy group $g$ are considered. $S_{g'g}$ is the
group integrated differential scattering coefficient. In local thermal
equilibrium the weighting function $I_{\nu'}$ is given by Planck's function
approximately. One therefore obtains for $S_{g'g}$

\begin{equation}\label{eqn.DiffScattLTE}
  S_{g'g} = \frac{1}{B_{g'}}\int_{g'} d\nu'\; \frac{B_{\nu'}}{\nu'} 
  \int_g d\nu\;\nu\Sigma_s(\nu'\rightarrow\nu,\mathbf{\Omega'}\cdot\mathbf{\Omega}).
\end{equation}

The Legendre expansion of $\displaystyle
\Sigma_{s}(\nu'\rightarrow\nu,\mathbf{\Omega'}\cdot\mathbf{\Omega})$ reads

\begin{equation}\label{eqn.DiffScattLegendreExpansion}
  \Sigma_{s}(\nu'\rightarrow\nu, \mathbf{\Omega'}\cdot\mathbf{\Omega}) = 
  \sum^{\infty}_{l=0}\frac{2l+1}{4\pi}\Sigma_{sl}(\nu'\rightarrow\nu)
  P_l(\mathbf{\Omega'}\cdot\mathbf{\Omega}).
\end{equation}

where the expansion coefficients $\Sigma_{sl}(\nu'\rightarrow\nu)$ are defined by

\begin{equation}\label{eqn.DiffScattLegendreExpansionCoefficient}
  \Sigma_{sl}(\nu'\rightarrow\nu) = 2\pi\int^1_{-1} d\mu_0\; 
  \Sigma_s(\nu'\rightarrow\nu,\mu_0)P_l(\mu_0)
\end{equation}

where $\displaystyle\mu_0 = \mathbf{\Omega'}\cdot\mathbf{\Omega}$ is the
scattering angle. By Using (\ref{eqn.DiffScattLegendreExpansion}) and
(\ref{eqn.DiffScattLegendreExpansionCoefficient}) in (\ref{eqn.DiffScattLTE})
one defines the Legendre ordered group integrated differential scattering
coefficient

\begin{equation}\label{eqn.MomentsDiffScattGeneral}
 S_{lg'g} = \frac{2\pi}{B_{g'}}\int^{-1}_1 d\mu_0 P_l(\mu_0)\int_{g'}d\nu'
 \frac{B_{\nu'}}{\nu'}\int_g d\nu \nu\Sigma_s(\nu'\rightarrow\nu, \mu_0).
\end{equation}

$l$ marks the Legendre order. By using the above transformations the
differential scattering contribution of the radiative transfer equation reads

\begin{equation}
  \mathcal{C}_g[I_{\nu}] = \sum^{\infty}_{l=0}\frac{2l+1}{4\pi}\int_{4\pi}d\mathbf{\Omega'}
 P_l(\mathbf{\Omega'\cdot\Omega})\sum^{gmax(g)}_{g'=gmin(g)}S_{lg'g}
I_{g'}(\mathbf{\Omega'}).
\end{equation}

In the next section the contributions $S_{lg'g}$ are evaluated. Beside
numerical calculations analytical expressions are derived.

\subsection{Moments of the group transfer scattering cross section}

The differential scattering moments are given by
(\ref{eqn.MomentsDiffScattGeneral}). For further purposes $S_{lg'g}$ is transformed
by help of (\ref{eqn.PlanckFunction}) and $\gamma = h\nu/m_e c^2$ in

\begin{equation}\label{eqn.MomentsDiffScattGeneralTransformed}
  S_{lg'g} = \frac{4\pi (m_e c^2)^3}{h^2 c^2}\frac{1}{B_{g'}}\int^1_{-1}d\mu_0 P_l(\mu_0)
  \int_{g'}\frac{d\gamma'\gamma'^2}{\left(\exp(\gamma'm_e c^2/\theta)-1\right)}
  \times\int_g d\nu \nu\Sigma_s(\nu'\rightarrow\nu,\mu_0).
\end{equation}


\subsubsection{Group transfer scattering cross section in the Thomson limit}

The Thomson differential cross section is given by \cite{Pomraning:73}

\begin{equation}
 \Sigma_s(\nu'\rightarrow\nu,\mu_0) = \frac{3}{16\pi}\Sigma_{Th}(1+\mu^2_0)\delta(\nu'-\nu).
\end{equation}

Using the above expression in (\ref{eqn.MomentsDiffScattGeneralTransformed}) yields

\begin{eqnarray}
  S_{lg'g} = \frac{2\pi}{B_{g'}}\int^1_{-1} d\mu_0 P_l(\mu_0)
  \int_{g'} d\nu' \frac{B_{\nu'}}{\nu'}
  \int_g d\nu \nu \frac{3}{16\pi} 
  \Sigma_{Th}(1 + \mu^2_0) \delta(\nu'-\nu)\nonumber
  = \frac{3}{8}\Sigma_{Th}\int^1_{-1} d\mu_0 P_l(\mu_0)(1 + \mu^2_0).
\end{eqnarray}

The Thomson cross section does not depend on frequency of the incident
photons. No frequency shifts occur during the scattering. Hence, the
out-scattered photons belong to the same energy group as the in-scattered
photons. Owing to the definition of the Legendre polynomials one calculates

\begin{equation}
  1 + \mu^2_0 = \frac{4}{3}P_0(\mu_0) + \frac{2}{3}P_2(\mu_0)
\end{equation}

and therefore

\begin{eqnarray}\label{eqn.MomentsThomsonLimit}
  S_{0gg} &=& \frac{\Sigma_{Th}}{2}\int^1_{-1}d\mu_0 P_0(\mu_0)P_0(\mu_0)
  = \Sigma_{Th}\nonumber\\
  S_{1gg} &=& 0\nonumber\\
  S_{2gg} &=& \frac{\Sigma_{Th}}{4}\int^1_{-1}d\mu_0 P_2(\mu_0)P_2(\mu_0) 
  = \frac{1}{10}\Sigma_{Th}
\end{eqnarray}

or in terms of the Thomson unit $S_{0gg} = 1$, $S_{1gg} = 0$, $S_{2gg} =
1/10$. The orthogonality relation 

\begin{equation*}
  \int^1_{-1}d\mu_0\;P_n(\mu_0) P_m(\mu_0) = \frac{2}{2n + 1}\delta_{nm}
\end{equation*}

has been used while evaluating the transfer cross section. All higher moments 
are zero.


\subsubsection{Frequency shift formula}

The shift of frequency of the photon scattering on an electron at rest is given by

\begin{equation}
 \triangle\lambda = \lambda - \lambda' = \frac{h}{m_e c}\left(1 - \mu_0\right).
\end{equation}

In that case $\lambda$ and $\lambda'$ denote the wavelength of the 
scattered and in-scattered photon, respectively ($\lambda\ge\lambda'$, 
$\triangle\lambda\ge 0$). Switching to the energy group notation, 
using the relation $\lambda = c/\nu$ and considering extremal cases only,
one obtains

\begin{eqnarray}
 \frac{1}{\nu_g}-\frac{1}{\nu'_{g,max}} & = & \frac{2h}{m_e c^2}\qquad \mu_0 = -1\\[2mm]
 \frac{1}{\nu_g}-\frac{1}{\nu'_{g,min}} & = & 0\qquad \mu_0 = 1.
\end{eqnarray}

Therefore one gains

\begin{eqnarray}\label{eqn.NuHighNuLow}
 \nu'_{max,g} &=& \frac{\nu_g}{1 - 2\gamma_g}\\[2mm]
 \nu'_{min,g} &=& \nu_{g},
\end{eqnarray}

where $\gamma_g$ is defined by $\gamma_g = h\nu_g/m_e c^2$. There is
no up-scattering for electrons at rest, hence $S_{lg'g} = 0$, if $g' >
g$. Table (\ref{tab.maxfre}) shows the highest frequencies
$\nu'_{max,g}$ of photons which are scattered into group
$g$. Comparing the values for $h\nu'_{max,g}$ and $h\nu_{g-1}$ one
recognises $h\nu'_{max,g} < h\nu_{g-1}$. This depends on the
  chosen energy groups. Therefore one has in-scattering
contributions from group $g-1$ only. $S_{lg'g}=0$, if $g' <
g-1$. Because of no up-scattering we only have to calculate
$S_{lg-1g}$ and $S_{lgg}$. All other contributions are zero. Pritzker
et al. \cite{Pritzker:76} take $\nu_{g+1}$ for $\nu'_{min,g}$. This is
inappropriate. Their usage leads to a double counting procedure for
in-scattering contributions of group $g-1$ into $g$ and $g$ into
$g$. The choice of $\nu_{g}$ for $\nu'_{min,g}$ guarantees that only
photons with frequencies greater or equal than $\nu_{g}$ will be
considered as in-scattering terms for group $g$. In-scattering
contributions from group $g$ into group $g$ will be considered
separately.

\begin{table}[htbp]
 \begin{center}
  \begin{tabular}{|Q{6mm}|Q{16mm}|Q{17mm}|Q{17mm}|Q{17mm}|Q{17mm}|}
   \hline
   g & $h\nu_{g-1}$ & $h\nu_{g}$ & $\gamma_g $ & $h\nu'_{max,g}$ & $h\nu'_{min,g}$\\
   \hline
   \hline
   1 & -          & 5.000e+02 & -          & -          & -\\
   \hline
   2 & 5.000e+02  & 6.400e+01 & 9.785e-01 & 8.539e+01 & 6.400e+01\\
   \hline
   3 & 6.400e+01  & 1.600e+01 & 1.252e-01 & 1.707e+01 & 1.600e+01\\
   \hline
   4 & 1.600e+01  & 4.000e+00 & 7.828e-03 & 4.064e+00 & 4.000e+00\\
   \hline
   5 & 4.000e+00  & 1.000e+00 & 1.957e-03 & 1.004e+00 & 1.000e+00\\
   \hline
   6 & 1.000e+00  & 2.500e-01 & 4.892e-04 & 2.502e-01 & 2.500e-01\\
   \hline
  \end{tabular}
  \caption{Highest and smallest photon energies $h\nu'$ from which photons are 
    able to scatter into group $g$. The photon energies are given in keV.}
  \label{tab.maxfre}
 \end{center}
\end{table}


\subsubsection{General case of the group transfer scattering cross section}

First of all general solution for arbitrary photons energies by help of
numerical methods are presented. Secondly, for small photon energies an
analytical solution is derived.\\[2mm]

The Compton differential cross section reads \cite{Pomraning:73}

\begin{equation*}
    \Sigma_s (\nu'\rightarrow\nu,\mu_0) = \frac{3}{16\pi}\Sigma_{Th}
    \frac{\left(1+\mu_0^2\right)}{\left(1+\gamma'\left(1-\mu_0\right)\right)^2}
    \left(1+\frac{\gamma'\left(1-\mu_0\right)^2}{\left(1+\mu_0^2\right)
        \left(1+\gamma'\left(1-\mu_0\right)\right)}\right)
    \delta\left(\nu - \nu'\left(\frac{1}{1+\gamma'\left(1-\mu_0\right)}\right)\right).
\end{equation*}

Factors of the same order of $\mu_0$ are grouped together. Hence, the Compton differential cross 
section reads

\begin{equation}\label{eqn.ComptonScattTransPritzker}
 \begin{split}
  \Sigma_s (\nu'\rightarrow\nu,\mu_0) &= 
  \frac{3}{16\pi}\Sigma_{Th}\frac{(1+\gamma'+\gamma'^2)+(-\gamma'-2\gamma'^2)\mu_0+
    (1+\gamma'^+\gamma'^2)
  \mu^2_0}{\left(1+\gamma'(1-\mu_0) \right)^3}\\
  &-\frac{\gamma'\mu^3_0}{\left(1+\gamma'(1-\mu_0) \right)^3}\times\delta\left(\nu-\frac{\nu'}{1 +
  \gamma'(1-\mu_0)}\right).
 \end{split}
\end{equation}

(\ref{eqn.ComptonScattTransPritzker}) is inserted into 
(\ref{eqn.MomentsDiffScattGeneralTransformed}). In this way one obtains a triple integral. 
Since we are asking for the scattering probability of photons with all possible 
frequencies $\nu'$ for all possible angles $\mu_0$ the integral over $d\nu$ 
is evaluated by means of the $\delta$-distribution. The resulting fraction 
on the right hand side is abbreviated by

\begin{equation}\label{eqn.PRratio}
  \frac{P(\mu_0,\gamma')}{R(\mu_0,\gamma')} =
  \frac{(1 + \gamma' + \gamma'^2) + (-\gamma' -
    2\gamma'^2)\mu_0 + (1 + \gamma' + \gamma'^2)\mu^2_0 - \gamma' \mu^3_0}
  {(\gamma'^{-1} + 1 - \mu_0)^4}.
\end{equation}

Hence, the moments of the Compton differential cross section read

\begin{equation}\label{eqn.ComptonTransferGeneral}
 S_{lg'g} =  \frac{3}{8}\frac{\Sigma_{Th}}{B_{g'}}\frac{2(m_ c^2)^4}{c^2 h^3}
 \Gamma\left(\gamma^{**},\gamma^*\right)
\end{equation}

where the results from the frequency shift formula (\ref{eqn.NuHighNuLow}) have been used. Here 
$\gamma^*=\gamma_g, \gamma^{**}=\gamma_{g+1}$ if $g' = g$ and 
$\gamma^*=\gamma_g/1-2\gamma_g, \gamma^{**}=\gamma_g$ if $g' = g-1$. 
$\Gamma\left(\gamma^{**},\gamma^*\right)$ is given by

\begin{eqnarray}
 \Gamma\left(\gamma^{**},\gamma^*\right) &=& \int^{\gamma^*}_{\gamma^{**}}\frac{d\gamma'}
{\left(\exp\left(\gamma'm_e c^2/\theta\right) - 1\right)\gamma'} F_l(\gamma'),\\[2mm]
 F_l(\gamma') &=& \int^1_{-1} d\mu_0\;\frac{P(\mu_0,\gamma')}{R(\mu_0,\gamma')} P_l(\mu_0)  = 
 \sum^4_{i=1}F^{(i)}_l(\gamma'),
\end{eqnarray}

where the integrals $F^{(i)}_l$ are defined by

\begin{eqnarray}
  F^{(1)}_l(\gamma') &=& \int^1_{-1} d\mu_0\; \frac{-\gamma'\mu^3_0}
  {(\gamma'^{-1} + 1 - \mu_0)^4} P_l(\mu_0)\label{eqn.Fl1}\\
  F^{(2)}_l(\gamma') &=& \int^1_{-1} d\mu_0\; \frac{(1 + \gamma' + \gamma'^2)\mu^2_0}
  {(\gamma'^{-1} + 1 - \mu_0)^4} P_l(\mu_0)\label{eqn.Fl2}\\
  F^{(3)}_l(\gamma') &=& \int^1_{-1} d\mu_0\; \frac{(-\gamma' - 2\gamma'^2)\mu_0}
  {(\gamma'^{-1} + 1 - \mu_0)^4} P_l(\mu_0)\label{eqn.Fl3}\\
  F^{(4)}_l(\gamma') &=& \int^1_{-1} d\mu_0\; \frac{(1 + \gamma' + \gamma'^2)}
  {(\gamma'^{-1} + 1 - \mu_0)^4} P_l(\mu_0)\label{eqn.Fl4}.
\end{eqnarray}

The integrals (\ref{eqn.Fl1}-\ref{eqn.Fl4}) are analytically
evaluable. Taking the Legendre polynomials of zeroth order $F^{(i)}_0$
reads

\begin{eqnarray*}
 F^{(1)}_0(\gamma') & = & -\gamma'\left(\frac{6\tilde{a}}{1 - \tilde{a}^2} + 
 \frac{6\tilde{a}^3}{(1 - \tilde{a}^2)^2}
 + \frac{2\tilde{a}^3}{3}\frac{(3\tilde{a}^2 + 1)}{(1 - \tilde{a}^2)^3}
   +\ln\frac{\left|1 + \tilde{a}\right|}{\left|\tilde{a} - 1\right|}\right),\\
 F^{(2)}_0(\gamma') & = & -(1 + \gamma' + \gamma'^2)\left(\frac{2}{1 - \tilde{a}^2} + 
 \frac{4\tilde{a}^2}{(1 - \tilde{a}^2)^2}
 + \frac{2\tilde{a}^2}{3}\frac{(3\tilde{a}^2 + 1)}{(1 - \tilde{a}^2)^3}\right),\\
 F^{(3)}_0(\gamma') & = & -(\gamma' + 2\gamma'^2)\left(\frac{2\tilde{a}}{1 - \tilde{a}^2}
 + \frac{2\tilde{a}}{3}\frac{(3\tilde{a}^2 + 1)}{(1 - \tilde{a}^2)^3}\right),\\
 F^{(4)}_0(\gamma') & = & -\frac{2}{3}\left(1 + \gamma' + 
 \gamma'^2\right)\frac{(3\tilde{a}^2 + 1)}{(1 - \tilde{a}^2)^3},
\end{eqnarray*}


$\tilde{a}\;=\;-1/\gamma'-1$. The integral functions $F^{(i)}_l$ of higher order in $l$ are 
evaluated in the same manner. $F^{(i)}_l$ evaluated in this way is inserted in 
(\ref{eqn.ComptonTransferGeneral}). The integration over $\gamma'$ in $S_{lg'g}$ is carried out 
numerically. The results are given in tables (\ref{Table.SixGroupPlanckMeanComptonZerothMoments}),
 (\ref{Table.SixGroupPlanckMeanComptonFirstMoments}) and 
(\ref{Table.SixGroupPlanckMeanComptonSecondMoments}) and are in agreement with the ones 
obtained by Pritzker et al. \cite{Pritzker:76}.

\subsubsection{An asymptotic analytical solution}

While $S_{lg'g}$ expressed by (\ref{eqn.MomentsDiffScattGeneralTransformed}) is evaluable in a 
numerical way only, one can perform analytic calculations assuming the small photon energy limit 
\mbox{$\gamma\ll 1$}. In that case the differential scattering cross section 
\mbox{$\Sigma_s (\nu'\rightarrow\nu,\mu_0)$} reads \cite{Pomraning:73}

\begin{equation}
  \begin{split}
    \Sigma_s (\nu'&\rightarrow\nu,\mu_0) = \frac{3}{16\pi}\Sigma_{Th}(1 +
    \mu^2_0)\left[1 - 2\gamma(1-\mu_0) +\right.\\
    &\left.\gamma^2\frac{(1-\mu_0)^2(4 + 3\mu^2_0)}{1 + \mu^2_0}\right]
    \times\delta\left(\nu' - \nu\left[1 - \gamma(1-\mu_0) + \gamma^2(1 - \mu_0)^2\right]\right).
  \end{split}
\end{equation}

This term is inserted into (\ref{eqn.MomentsDiffScattGeneralTransformed}). 
The resulting triple integral is evaluated over $d\nu$ by means of the
$\delta$-distribution. In this way one obtains

\begin{equation}
 \begin{split}
  S_{lg'g} &= \frac{3}{8}\frac{\Sigma_{Th}}{B_{g'}}\frac{2(m_e c^2)^4}{c^2
  h^3}\int^1_{-1}d\mu_0 P_l(\mu_0)
  \int^{\gamma^*}_{\gamma^{**}}\frac{d\gamma'\gamma'^3}{\left(\exp(\gamma'm_e
  c^2/\theta)-1\right)}\times\left\lbrace 
  \left(1-3\gamma'+7\gamma'^2-6\gamma'^3+4\gamma'^4\right) +\right.\\
  &+(3\gamma'-14\gamma'^2+18\gamma'^3-16\gamma'^4)\mu_0
  +\left(1-3\gamma'+13\gamma'^2-23\gamma'^3+27\gamma'^4\right)\mu^2_0+\\
  &\left.+\left(3\gamma'-12\gamma'^2+21\gamma'^3-28\gamma'^4\right)\mu^3_0+
  +\left(6\gamma'^2-15\gamma'^3+22\gamma'^4\right)\mu^4_0+
  +\left(5\gamma'^3-12\gamma'^4\right)\mu^5_0+ 3\gamma'^4\mu^6_0\right\rbrace.
 \end{split}
\end{equation}

In lowest Legendre order, $P_0(\mu_0)$ = 1, the above expression
integrated over $\mu_0$ leads to

\begin{equation}\label{eqn.ComptonTransferZerothOrderAnalytic}
 \begin{split}
 S_{0g'g} &= \frac{3}{8}\frac{\Sigma_{Th}}{B_{g'}}\frac{2\theta^4}{c^2 h^3}\left\lbrace
 \frac{8}{3}\mathcal{I}_3(u^{**},u^*)-8\alpha\mathcal{I}_4(u^{**},u^*)
 +\frac{376}{15}\alpha^2\mathcal{I}_5(u^{**},u^*)\right.\\
 &-\left.\frac{100}{3}\alpha^3\mathcal{I}_6(u^{**},u^*) + 
\frac{1248}{35}\alpha^4\mathcal{I}_7(u^{**},u^*)\right\rbrace 
 \end{split}
\end{equation}

where $\displaystyle \alpha=\theta/m_e c^2$ and $\displaystyle
u^*=\gamma^*/\alpha, u^{**}=\gamma^{**}/\alpha$. The terms
$\mathcal{I}_k$ are derived in the appendix. The result is presented
in table (\ref{Table.SixGroupPlanckMeanComptonMomentsAnalyticZerothMoment}).
Proceeding with the next order group moment coefficient,
$P_1(\mu_0)$=$\mu_0$, yields

\begin{equation}\label{eqn.ComptonTransferFirstOrderAnalytic}
 \begin{split}
 S_{1g'g} = \frac{3}{8}\frac{\Sigma_{Th}}{B_{g'}}\frac{2\theta^4}{c^2 h^3}\left\lbrace
 \frac{16}{5}\alpha\mathcal{I}_4(u^{**},u^*)-\frac{212}{15}\alpha^2\mathcal{I}_5(u^{**},u^*)
 +\frac{764}{35}\alpha^3\mathcal{I}_6(u^{**},u^*) - 
\frac{236}{5}\alpha^4\mathcal{I}_7(u^{**},u^*)\right\rbrace.
 \end{split}
\end{equation}

The result is presented in table
(\ref{Table.SixGroupPlanckMeanComptonMomentsAnalyticFirstMoment}).
Finally, the second order group moment, $P_2(\mu_0)=0.5(-1+3\mu^2_0)$,
is considered.

\begin{equation}\label{eqn.ComptonTransferSecondOrderAnalytic}
 \begin{split}
  S_{2g'g} &= -\frac{1}{2}S_{0g'g} +\frac{9}{16}\frac{\Sigma_{Th}}{B_{g'}}\frac{2\theta^4}{c^2
 h^3}\left\lbrace
  \frac{16}{15}\mathcal{I}_3(u^{**},u^*)
  -\frac{16}{5}\alpha\mathcal{I}_4(u^{**},u^*)\right.\\
  &\left.+\frac{1216}{105}\alpha^2\mathcal{I}_5(u^{**},u^*)
  -\frac{612}{35}\alpha^3\mathcal{I}_6(u^{**},u^*)
  +\frac{2018}{105}\alpha^4\mathcal{I}_7(u^{**},u^*)\right\rbrace.
 \end{split}
\end{equation}

The result is presented in table
(\ref{Table.SixGroupPlanckMeanComptonMomentsAnalyticSecondMoment}).
One can see by the tables
(\ref{Table.SixGroupPlanckMeanComptonMomentsZerothMomentDeviation}),
(\ref{Table.SixGroupPlanckMeanComptonMomentsFirstMomentDeviation}) and
(\ref{Table.SixGroupPlanckMeanComptonMomentsSecondMomentDeviation})
the deviation from the numerically evaluated results is very small
except for the highest energy group. Comparing to the published
results from Pritzker one obtains the same results without much
numerical effort.

\section{Conclusion and discussion}

In this paper, numerical and analytical expressions for the
multi-group total scattering and transfer scattering cross section are
found. Scattering is analysed by Thomson and Compton scattering for
free electrons at rest. Analytical terms have been obtained in the
small photon energy limit. From tables
(\ref{Table.SixGroupPlanckMeanComptonMomentsZerothMomentDeviation}),
(\ref{Table.SixGroupPlanckMeanComptonMomentsFirstMomentDeviation}) and
(\ref{Table.SixGroupPlanckMeanComptonMomentsSecondMomentDeviation})
one can see that the deviation from the general valid results, shown
in tables (\ref{Table.SixGroupPlanckMeanComptonZerothMoments}),
(\ref{Table.SixGroupPlanckMeanComptonFirstMoments}) and
(\ref{Table.SixGroupPlanckMeanComptonSecondMoments}), is a few
percents only.  The Compton cross section starts to deviate from the
Thomson limit at high temperatures and high energy groups, e.g. tables
(\ref{Table.SixGroupPlanckMeanComptonCS}) and
(\ref{Table.SixGroupPlanckMeanComptonZerothMoments}).  For
temperatures below 3 keV and energy groups with appreciable
contributions to the Planck spectrum these deviations are small. The
moment $l=1$, which vanishes in the Thomson limit, should be taken
into account as like as the transfer cross sections $S_{lg-1g}$ which
become comparable to $S_{lgg}$ for temperatures above 1 keV. As
pointed out by Pritzker et al. \cite{Pritzker:75} the diffusion
technique is an adequate approximation describing the radiative
processes in a plasma at temperatures below 3 keV. In that case the
plasma is optical thick and scattering is not important. For higher
temperatures the plasma becomes optical thin. In such regions Compton
scattering should be considered for a better approximation of the ERT.

\section{Acknowledgement}

I am grateful to E. Lieder and R. K\"{u}lheim giving corrections and
comments.

\section*{Appendix}

The solution of integrals of the type

\begin{equation}
  \mathcal{I}_k(a,b) = \frac{1}{k!}\int^b_a \frac{du u^k}{\exp(u) - 1}
\end{equation}

taking $a=0$ and $b=\infty$ is given by the Zeta-function

\begin{equation}
  \zeta(k) = \sum^{\infty}_{n=1} \frac{1}{n^{k+1}}.
\end{equation}

Such integrals appear in one-energy group calculations. More effort is required
setting $a$ and $b$ to arbitrary real values. A lengthy technical integration
procedure yields to

\begin{equation}\label{eqn.SpecialInt}
 \mathcal{I}_k(a,b) =
 \sum^{\infty}_{n=1}\frac{1}{n^{k+1}}\left[\exp(-\tilde{a})\left(\sum^k_{i=0}\frac{k!}{i!}
 \tilde{a}^i \right)  - \exp(-\tilde{b})\left(\sum^k_{i=0}\frac{k!}{i!}\tilde{b}^i \right)\right],
\end{equation}

where $\tilde{a}=na$ and $\tilde{b}=nb$.

\bibliographystyle{apsrev4-1}
\bibliography{Ref}

\newpage

\vspace*{\fill}
\begin{table}[htbp]
  \begin{center}
    \begin{tabular}{|Q{8mm}|Q{1.5cm}|Q{1.5cm}|Q{1.5cm}|Q{1.5cm}|Q{1.5cm}|}
      \hline
      g/$\theta$ & 1.00e-01 & 3.16e-01 & 1.00e+00 & 3.16e+00 & 1.00e+01\\
      \hline
      \hline
      1 & 8.140e-01 & 8.136e-01 & 8.121e-01 & 8.067e-01 & 7.851e-01 
      \\
      \hline
      2 & 9.420e-01 & 9.413e-01 & 9.385e-01 & 9.260e-01 & 8.831e-01 
      \\
      \hline
      3 & 9.843e-01 & 9.832e-01 & 9.781e-01 & 9.635e-01 & 9.574e-01 
      \\
      \hline
      4 & 9.956e-01 & 9.937e-01 & 9.899e-01 & 9.887e-01 & 9.884e-01 
      \\
      \hline
      5 & 9.982e-01 & 9.973e-01 & 9.971e-01 & 9.971e-01 & 9.970e-01 
      \\
      \hline
      6 & 9.996e-01 & 9.994e-01 & 9.994e-01 & 9.994e-01 & 9.994e-01 
      \\
      \hline
    \end{tabular}
    \caption[Six-group Planck mean Compton cross section]{Six-group Planck mean 
      Compton cross section in Thomson units evaluated at different temperatures 
      $\theta$ in keV.}
    \label{Table.SixGroupPlanckMeanComptonCS}
  \end{center}
  
  \begin{center}
    \begin{tabular}{|Q{8mm}|Q{1.5cm}|Q{1.5cm}|Q{1.5cm}|Q{1.5cm}|Q{1.5cm}|}
      \hline
      g/$\theta$ & 1.00e-01 & 3.16e-01 & 1.00e+00 & 3.16e+00 & 1.00e+01\\
      \hline
      \hline
      1 & 8.309e-01 & 8.306e-01 & 8.297e-01 & 8.266e-01 & 8.191e-01\\
      \hline
      2 & 9.421e-01 & 9.414e-01 & 9.386e-01 & 9.265e-01 & 8.871e-01\\
      \hline
      3 & 9.843e-01 & 9.832e-01 & 9.780e-01 & 9.635e-01 & 9.574e-01\\
      \hline
      4 & 9.956e-01 & 9.937e-01 & 9.899e-01 & 9.887e-01 & 9.884e-01\\
      \hline
      5 & 9.982e-01 & 9.973e-01 & 9.971e-01 & 9.971e-01 & 9.970e-01\\
      \hline
      6 & 9.996e-01 & 9.994e-01 & 9.994e-01 & 9.994e-01 & 9.994e-01\\
      \hline
    \end{tabular}
    \caption[Analytical evaluated Six-group Planck mean Compton cross section in 
      small photon energy limit.]{The table shows the six-group Planck mean Compton
      cross section in Thomson units evaluated at different temperatures $\theta$ in 
      keV. The constants $\Sigma_{stg}/\Sigma_{Th}$ obtained from 
      (\ref{eqn.ScatteringSigmaSTGAnalytic}) are independent from material and  
      density. The results are in good agreement with those in table 
      (\ref{Table.SixGroupPlanckMeanComptonCS}) except for the case, where $\gamma\ll 1$ 
      is not valid.}
    \label{Table.SixGroupPlanckMeanComptonCSAnalytic}
  \end{center}
  
  \begin{center}
    \begin{tabular}{|Q{8mm}|Q{1.5cm}|Q{1.5cm}|Q{1.5cm}|Q{1.5cm}|Q{1.5cm}|}
      \hline
      g/$\theta$ & 1.00e-01 & 3.16e-01 & 1.00e+00 & 3.16e+00 & 1.00e+01\\
      \hline
      \hline
      1 &  2.077 &  2.095 &  2.172 &  2.474 &  4.322 
      \\
      \hline
      2 &  0.010 &  0.011 &  0.016 &  0.053 &  0.456 
      \\
      \hline
      3 &  0.000 &  0.000 &  0.001 &  0.000 &  0.001 
      \\
      \hline
      4 &  0.000 &  0.000 &  0.000 &  0.000 &  0.000 
      \\
      \hline
      5 &  0.000 &  0.000 &  0.000 &  0.000 &  0.000 
      \\
      \hline
      6 &  0.000 &  0.000 &  0.000 &  0.000 &  0.000 
      \\
      \hline
    \end{tabular}
    \caption[Accuracy of the multi-group Compton cross section in small photon 
      energy limit.]{Accuracy of the multi-group Compton cross section in small 
      photon energy limit. The table shows the deviation in percent from the 
      general valid multi-group term (\ref{eqn.Stg}). Precise agreement between the 
      general case (table (\ref{Table.SixGroupPlanckMeanComptonCSAnalytic})) and 
      small energy limit (table (\ref{Table.SixGroupPlanckMeanComptonCS})) are 
      obtained for the energy groups two to five up to temperatures of around 4 keV.}
    \label{Table.SixGroupPlanckMeanComptonCSDeviation}
  \end{center}
\end{table}
\vspace*{\fill}

\newpage

\vspace*{\fill}
\begin{table}[htbp]
  \begin{center}
    \begin{tabular}{|Q{1cm}|Q{1.4cm}|Q{1.4cm}|Q{1.4cm}|Q{1.4cm}|Q{1.4cm}|}
      \hline
      g',g/$\theta$ & 1.00e-01 & 3.16e-01 & 1.00e+00 & 3.16e+00 & 1.00e+01\\
      \hline
      \hline
      0,1 & - & - & - & - & -\\
      1,1 & 7.309e-01 & 7.302e-01 & 7.281e-01 & 7.206e-01 & 6.910e-01\\
      \hline
      1,2 & 7.309e-01 & 7.302e-01 & 7.281e-01 & 7.189e-01 & 5.329e-01\\
      2,2 & 9.140e-01 & 9.129e-01 & 9.087e-01 & 8.906e-01 & 8.287e-01\\
      \hline
      2,3 & 9.140e-01 & 8.756e-01 & 5.366e-01 & 1.540e-01 & 2.059e-02\\
      3,3 & 9.765e-01 & 9.749e-01 & 9.672e-01 & 9.457e-01 & 9.366e-01\\
      \hline
      3,4 & 4.354e-01 & 1.427e-01 & 2.812e-02 & 3.481e-03 & 1.176e-03\\
      4,4 & 9.934e-01 & 9.906e-01 & 9.849e-01 & 9.831e-01 & 9.827e-01\\
      \hline
      4,5 & 2.821e-02 & 4.664e-03 & 6.242e-04 & 2.647e-04 & 2.073e-04\\
      5,5 & 9.973e-01 & 9.960e-01 & 9.957e-01 & 9.956e-01 & 9.956e-01\\
      \hline
      5,6 & 7.448e-04 & 1.194e-04 & 6.118e-05 & 5.062e-05 & 4.783e-05\\
      6,6 & 9.992e-01 & 9.991e-01 & 9.990e-01 & 9.990e-01 & 9.990e-01\\
      \hline
    \end{tabular}
    \caption[Zeroth order moments of the six-group Planck weighted Compton 
      scattering transfer cross section.]{Zeroth order moments of the six-group 
      Planck weighted Compton scattering transfer cross section. The results are 
      given in units of the Thomson cross section $S_{lg'g}/\Sigma_{Th}$. The
      temperature $\theta$ is given in units of keV.}
    \label{Table.SixGroupPlanckMeanComptonZerothMoments}
  \end{center}
  
  \begin{center}
    \begin{tabular}{|Q{1cm}|Q{1.4cm}|Q{1.4cm}|Q{1.4cm}|Q{1.4cm}|Q{1.4cm}|}
      \hline
      g',g/$\theta$ & 1.00e-01 & 3.16e-01 & 1.00e+00 & 3.16e+00 & 1.00e+01\\
      \hline
      \hline
      0,1 & - & - & - & - & -\\
      1,1 & 7.502e-01 & 7.497e-01 & 7.482e-01 & 7.430e-01 & 7.264e-01 \\
      \hline
      1,2 & 7.502e-01 & 7.497e-01 & 7.482e-01 & 7.411e-01 & 5.533e-01 \\
      2,2 & 9.144e-01 & 9.133e-01 & 9.093e-01 & 8.916e-01 & 8.341e-01 \\
      \hline
      2,3 & 9.144e-01 & 8.760e-01 & 5.369e-01 & 1.541e-01 & 2.060e-02 \\
      3,3 & 9.765e-01 & 9.749e-01 & 9.673e-01 & 9.459e-01 & 9.368e-01 \\
      \hline
      3,4 & 4.354e-01 & 1.427e-01 & 2.812e-02 & 3.481e-03 & 1.176e-03 \\
      4,4 & 9.934e-01 & 9.906e-01 & 9.849e-01 & 9.831e-01 & 9.827e-01 \\
      \hline
      4,5 & 2.821e-02 & 4.664e-03 & 6.242e-04 & 2.647e-04 & 2.073e-04 \\
      5,5 & 9.973e-01 & 9.960e-01 & 9.957e-01 & 9.956e-01 & 9.956e-01 \\
      \hline
      5,6 & 7.448e-04 & 1.194e-04 & 6.118e-05 & 5.062e-05 & 4.783e-05 \\
      6,6 & 9.992e-01 & 9.991e-01 & 9.990e-01 & 9.990e-01 & 9.990e-01 \\
      \hline
    \end{tabular}
    \caption[Analytical evaluated Six-group Planck weighted moments of zeroth 
      Legendre order of the Compton  scattering transfer cross section in the low 
      photon limit]{Zeroth order moment of the six-group Planck weighted Compton 
      scattering transfer cross section in the case of small photon energies. The 
      results are given in units of the Thomson cross section $S_{lg'g}/\Sigma_{Th}$. 
      The approximation fails at high photon energy groups and high temperatures. 
      In all other cases the results are in good agreement with those obtained by 
      numerical integration. Refer to table (\ref{Table.SixGroupPlanckMeanComptonZerothMoments}) 
      additionally.}
    \label{Table.SixGroupPlanckMeanComptonMomentsAnalyticZerothMoment}
  \end{center}
  
  \begin{center}
    \begin{tabular}{|Q{1cm}|Q{1.4cm}|Q{1.4cm}|Q{1.4cm}|Q{1.4cm}|Q{1.4cm}|}
      \hline
      g',g/$\theta$ & 1.00e-01 & 3.16e-01 & 1.00e+00 & 3.16e+00 & 1.00e+01\\
      \hline
      \hline
      0,1 & - & - & - & - & -\\
      1,1 &  2.651 &  2.673 &  2.761 &  3.106 &  5.120 \\
      \hline
      1,2 &  2.646 &  2.673 &  2.761 &  3.097 &  3.813 \\
      2,2 &  0.047 &  0.049 &  0.057 &  0.119 &  0.652 \\
      \hline
      2,3 &  0.047 &  0.048 &  0.050 &  0.050 &  0.051 \\
      3,3 &  0.001 &  0.001 &  0.003 &  0.014 &  0.021 \\
      \hline
      3,4 &  0.001 &  0.001 &  0.001 &  0.001 &  0.001 \\
      4,4 &  0.000 &  0.000 &  0.000 &  0.000 &  0.000 \\
      \hline
      4,5 &  0.000 &  0.000 &  0.000 &  0.000 &  0.000 \\
      5,5 &  0.000 &  0.000 &  0.000 &  0.000 &  0.000 \\
      \hline
      5,6 &  0.000 &  0.000 &  0.000 &  0.000 &  0.000 \\
      6,6 &  0.000 &  0.000 &  0.000 &  0.000 &  0.000 \\
      \hline
    \end{tabular}
    \caption[Accuracy of the multi-group Compton transfer cross section to zeroth 
      order in the small photon energy limit.]{Accuracy of the multi-group Compton 
      transfer cross section to zeroth order in the small photon energy limit. The 
      deviation is given in percent comparing the general valid 
      term  (\ref{eqn.ComptonTransferGeneral}) with small photon energy approximation
      (\ref{eqn.ComptonTransferZerothOrderAnalytic}). The results are in agreement to 
      those achieved by the general valid expression (\ref{eqn.ComptonTransferGeneral}) 
      except for the highest energy groups.}
    \label{Table.SixGroupPlanckMeanComptonMomentsZerothMomentDeviation}
  \end{center}
\end{table}
\vspace*{\fill}

\newpage

\vspace*{\fill}
\begin{table}[htbp]
  \begin{center}
    \begin{tabular}{|Q{1cm}|Q{1.4cm}|Q{1.4cm}|Q{1.4cm}|Q{1.4cm}|Q{1.4cm}|}
      \hline
      g',g/$\theta$ & 1.00e-01 & 3.16e-01 & 1.00e+00 & 3.16e+00 & 1.00e+01\\
      \hline
      \hline
      0,1 & - & - & - & - & -\\
      1,1 & 9.291e-02 & 9.310e-02 & 9.370e-02 & 9.573e-02 & 1.032e-01\\
      \hline
      1,2 & 9.291e-02 & 9.310e-02 & 9.370e-02 & 9.544e-02 & 7.486e-02\\
      2,2 & 3.307e-02 & 3.348e-02 & 3.499e-02 & 4.148e-02 & 6.237e-02\\
      \hline
      2,3 & 3.307e-02 & 3.203e-02 & 1.982e-02 & 5.711e-03 & 7.642e-04\\
      3,3 & 9.313e-03 & 9.938e-03 & 1.290e-02 & 2.114e-02 & 2.459e-02\\
      \hline
      3,4 & 4.074e-03 & 1.336e-03 & 2.633e-04 & 3.260e-05 & 1.101e-05\\
      4,4 & 2.632e-03 & 3.742e-03 & 5.995e-03 & 6.712e-03 & 6.873e-03\\
      \hline
      4,5 & 6.619e-05 & 1.094e-05 & 1.465e-06 & 6.212e-07 & 4.865e-07\\
      5,5 & 1.072e-03 & 1.591e-03 & 1.726e-03 & 1.757e-03 & 1.766e-03\\
      \hline
      5,6 & 4.372e-07 & 7.010e-08 & 3.591e-08 & 2.971e-08 & 2.807e-08\\
      6,6 & 4.010e-04 & 4.298e-04 & 4.368e-04 & 4.387e-04 & 4.393e-04\\
      \hline
    \end{tabular}
    \caption[First order moments of the six-group Planck weighted Compton 
      scattering transfer cross section.]{First order moments of the six-group 
      Planck weighted Compton scattering transfer cross section. The results are given 
      in units of the Thomson cross section $S_{lg'g}/\Sigma_{Th}$. The 
      temperature $\theta$ is given in units of keV.}
    \label{Table.SixGroupPlanckMeanComptonFirstMoments}
  \end{center}
  
  \begin{center}
    \begin{tabular}{|Q{1cm}|Q{1.4cm}|Q{1.4cm}|Q{1.4cm}|Q{1.4cm}|Q{1.4cm}|}
      \hline
      g',g/$\theta$ & 1.00e-01 & 3.16e-01 & 1.00e+00 & 3.16e+00 & 1.00e+01\\
      \hline
      \hline
      0,1 & - & - & - & - & -\\
      1,1 & 7.891e-02 & 7.895e-02 & 7.909e-02 & 7.933e-02 & 7.566e-02 \\
      \hline
      1,2 & 7.891e-02 & 7.895e-02 & 7.909e-02 & 7.914e-02 & 5.972e-02 \\
      2,2 & 3.279e-02 & 3.319e-02 & 3.465e-02 & 4.078e-02 & 5.861e-02 \\
      \hline
      2,3 & 3.279e-02 & 3.175e-02 & 1.965e-02 & 5.660e-03 & 7.574e-04 \\
      3,3 & 9.308e-03 & 9.932e-03 & 1.289e-02 & 2.105e-02 & 2.446e-02 \\
      \hline
      3,4 & 4.072e-03 & 1.335e-03 & 2.631e-04 & 3.258e-05 & 1.101e-05 \\
      4,4 & 2.632e-03 & 3.742e-03 & 5.993e-03 & 6.710e-03 & 6.871e-03 \\
      \hline
      4,5 & 6.618e-05 & 1.094e-05 & 1.465e-06 & 6.212e-07 & 4.865e-07 \\
      5,5 & 1.072e-03 & 1.591e-03 & 1.726e-03 & 1.757e-03 & 1.766e-03 \\
      \hline
      5,6 & 4.372e-07 & 7.010e-08 & 3.591e-08 & 2.971e-08 & 2.813e-08 \\
      6,6 & 4.010e-04 & 4.298e-04 & 4.368e-04 & 4.387e-04 & 4.393e-04 \\
      \hline
    \end{tabular}
    \caption[Analytical evaluated Six-group Planck weighted moments of first Legendre 
      order of the Compton scattering transfer cross section in low photon limit]{First 
      order moments of the six-group Planck weighted Compton scattering transfer 
      cross section for the case of small photon energies. The results are given in 
      units of the Thomson cross section $S_{lg'g}/\Sigma_{Th}$. The approximation 
      fails at high photon energy groups and high temperatures. In all other cases the 
      results are in very good agreement with those obtained by numerical integration. 
      Refer to table (\ref{Table.SixGroupPlanckMeanComptonFirstMoments}) additionally.}
    \label{Table.SixGroupPlanckMeanComptonMomentsAnalyticFirstMoment}
  \end{center}

  \begin{center}
    \begin{tabular}{|Q{1cm}|Q{1.4cm}|Q{1.4cm}|Q{1.4cm}|Q{1.4cm}|Q{1.4cm}|}
      \hline
      g',g/$\theta$ & 1.00e-01 & 3.16e-01 & 1.00e+00 & 3.16e+00 & 1.00e+01\\
      \hline
      \hline
      0,1 & - & - & - & - & -\\
      1,1 & 15.072 & 15.192 & 15.584 & 17.129 & 26.703 \\
      \hline
      1,2 & 15.076 & 15.192 & 15.584 & 17.084 & 20.218 \\
      2,2 &  0.846 &  0.872 &  0.980 &  1.702 &  6.019 \\
      \hline
      2,3 &  0.846 &  0.867 &  0.885 &  0.892 &  0.894 \\
      3,3 &  0.054 &  0.063 &  0.134 &  0.412 &  0.521 \\
      \hline
      3,4 &  0.052 &  0.052 &  0.052 &  0.052 &  0.052 \\
      4,4 &  0.004 &  0.011 &  0.028 &  0.033 &  0.034 \\
      \hline
      4,5 &  0.003 &  0.003 &  0.003 &  0.003 &  0.003 \\
      5,5 &  0.001 &  0.002 &  0.002 &  0.002 &  0.002 \\
      \hline
      5,6 &  0.000 &  0.000 &  0.000 &  0.000 &  0.207 \\
      6,6 &  0.000 &  0.000 &  0.000 &  0.000 &  0.000 \\
      \hline
    \end{tabular}
    \caption[Accuracy of the multi-group Compton transfer cross section to first
      order in the small photon energy limit.]{Accuracy of the multi-group Compton
      transfer cross section to first order in the small photon energy limit. The 
      deviation in percent comparing the general valid term (\ref{eqn.ComptonTransferGeneral})
      with the small photon energy approximation (\ref{eqn.ComptonTransferFirstOrderAnalytic}). 
      The results are in very good agreement to those achieved by the general valid 
      expression (\ref{eqn.ComptonTransferGeneral}) except for the highest energy groups.}
    \label{Table.SixGroupPlanckMeanComptonMomentsFirstMomentDeviation}
  \end{center}
\end{table}
\vspace*{\fill}

\newpage

\vspace*{\fill}
\begin{table}[htbp]
  \begin{center}
    \begin{tabular}{|Q{1cm}|Q{1.4cm}|Q{1.4cm}|Q{1.4cm}|Q{1.4cm}|Q{1.4cm}|}
      \hline
      g',g/$\theta$ & 1.00e-01 & 3.16e-01 & 1.00e+00 & 3.16e+00 & 1.00e+01\\
      \hline
      \hline
      0,1 & - & - & - & - & -\\
      1,1 & 8.041e-02 & 8.039e-02 & 8.029e-02 & 7.997e-02 & 7.890e-02\\
      \hline
      1,2 & 8.041e-02 & 8.039e-02 & 8.029e-02 & 7.977e-02 & 5.987e-02\\
      2,2 & 9.213e-02 & 9.204e-02 & 9.170e-02 & 9.029e-02 & 8.605e-02\\
      \hline
      2,3 & 9.213e-02 & 8.827e-02 & 5.411e-02 & 1.553e-02 & 2.076e-03\\
      3,3 & 9.770e-02 & 9.755e-02 & 9.684e-02 & 9.489e-02 & 9.408e-02\\
      \hline
      3,4 & 4.356e-02 & 1.428e-02 & 2.814e-03 & 3.483e-04 & 1.177e-04\\
      4,4 & 9.934e-02 & 9.907e-02 & 9.851e-02 & 9.834e-02 & 9.830e-02\\
      \hline
      4,5 & 2.821e-03 & 4.664e-04 & 6.242e-05 & 2.647e-05 & 2.073e-05\\
      5,5 & 9.973e-02 & 9.960e-02 & 9.957e-02 & 9.956e-02 & 9.956e-02\\
      \hline
      5,6 & 7.448e-05 & 1.194e-05 & 6.118e-06 & 5.062e-06 & 4.783e-06\\
      6,6 & 9.992e-02 & 9.991e-02 & 9.990e-02 & 9.990e-02 & 9.990e-02\\
      \hline
    \end{tabular}
    \caption[Second order moments of the six-group Planck weighted Compton 
      scattering transfer cross section.]{Second order moments of the six-group 
      Planck weighted Compton scattering transfer cross section. The results are 
      given in units of the Thomson cross section $S_{lg'g}/\Sigma_{Th}$. 
      The temperature $\theta$ is given in units of keV.}
    \label{Table.SixGroupPlanckMeanComptonSecondMoments}
  \end{center}
  
  \begin{center}
    \begin{tabular}{|Q{1cm}|Q{1.4cm}|Q{1.4cm}|Q{1.4cm}|Q{1.4cm}|Q{1.4cm}|}
      \hline
      g',g/$\theta$ & 1.00e-01 & 3.16e-01 & 1.00e+00 & 3.16e+00 & 1.00e+01\\
      \hline
      \hline
      0,1 & - & - & - & - & -\\
      1,1 & 8.486e-02 & 8.487e-02 & 8.490e-02 & 8.507e-02 & 8.665e-02 \\
      \hline
      1,2 & 8.486e-02 & 8.487e-02 & 8.490e-02 & 8.484e-02 & 6.443e-02 \\
      2,2 & 9.224e-02 & 9.215e-02 & 9.184e-02 & 9.056e-02 & 8.735e-02 \\
      \hline
      2,3 & 9.224e-02 & 8.838e-02 & 5.418e-02 & 1.555e-02 & 2.079e-03 \\
      3,3 & 9.770e-02 & 9.755e-02 & 9.684e-02 & 9.492e-02 & 9.413e-02 \\
      \hline
      3,4 & 4.357e-02 & 1.428e-02 & 2.814e-03 & 3.483e-04 & 1.177e-04 \\
      4,4 & 9.934e-02 & 9.907e-02 & 9.852e-02 & 9.834e-02 & 9.830e-02 \\
      \hline
      4,5 & 2.821e-03 & 4.664e-04 & 6.242e-05 & 2.647e-05 & 2.073e-05 \\
      5,5 & 9.973e-02 & 9.960e-02 & 9.957e-02 & 9.956e-02 & 9.956e-02 \\
      \hline
      5,6 & 7.448e-05 & 1.194e-05 & 6.118e-06 & 5.062e-06 & 4.783e-06 \\
      6,6 & 9.992e-02 & 9.991e-02 & 9.990e-02 & 9.990e-02 & 9.990e-02 \\
      \hline
    \end{tabular}
    \caption[Analytical evaluated Six-group Planck weighted moments of second 
      Legendre order to the Compton scattering transfer cross section in low photon 
      limit]{Second order moments of the six-group Planck weighted Compton scattering
      transfer cross section in the case of small photon energies. The results are 
      given in units of the Thomson cross section $S_{lg'g}/\Sigma_{Th}$. The 
      approximation fails at high photon energy groups and high temperatures. In 
      all other cases the results are in very good agreement with those obtained by 
      numerical integration. Refer to table 
      (\ref{Table.SixGroupPlanckMeanComptonSecondMoments}) additionally.}
    \label{Table.SixGroupPlanckMeanComptonMomentsAnalyticSecondMoment}
  \end{center}
  
  \begin{center}
    \begin{tabular}{|Q{1cm}|Q{1.4cm}|Q{1.4cm}|Q{1.4cm}|Q{1.4cm}|Q{1.4cm}|}
      \hline
      g',g/$\theta$ & 1.00e-01 & 3.16e-01 & 1.00e+00 & 3.16e+00 & 1.00e+01\\
      \hline
      \hline
      0,1 & - & - & - & - & -\\
      1,1 &  5.533 &  5.577 &  5.742 &  6.371 &  9.824 \\
      \hline
      1,2 &  5.528 &  5.577 &  5.742 &  6.355 &  7.631 \\
      2,2 &  0.120 &  0.125 &  0.147 &  0.298 &  1.505 \\
      \hline
      2,3 &  0.120 &  0.124 &  0.128 &  0.129 &  0.130 \\
      3,3 &  0.002 &  0.003 &  0.007 &  0.037 &  0.054 \\
      \hline
      3,4 &  0.002 &  0.002 &  0.002 &  0.002 &  0.002 \\
      4,4 &  0.000 &  0.000 &  0.001 &  0.001 &  0.001 \\
      \hline
      4,5 &  0.000 &  0.000 &  0.000 &  0.000 &  0.000 \\
      5,5 &  0.000 &  0.000 &  0.000 &  0.000 &  0.000 \\
      \hline
      5,6 &  0.000 &  0.000 &  0.000 &  0.000 &  0.000 \\
      6,6 &  0.000 &  0.000 &  0.000 &  0.000 &  0.000 \\
      \hline
    \end{tabular}
    \caption[Accuracy of the multi-group Compton transfer cross section of 
      second order in the small photon energy limit.]{Accuracy of the multi-group 
      Compton transfer cross section of second order in small photon energy limit.
      The deviation is given in percent comparing the general valid term
      (\ref{eqn.ComptonTransferGeneral}) with the small photon energy approximation
      (\ref{eqn.ComptonTransferSecondOrderAnalytic}). The results are in very good 
      agreement with those achieved by the general valid expression 
      (\ref{eqn.ComptonTransferGeneral}) except for the highest energy groups.}
    \label{Table.SixGroupPlanckMeanComptonMomentsSecondMomentDeviation}
  \end{center}
\end{table}
\vspace*{\fill}

\end{document}